# Institutional Learning and Volatility Transmission in ASEAN Equity Markets: A Network-Integrated Regime-Dependent Approach

*corresponding author:Junlin Yang*

*affliction:Wenzhou-Kean Unversity*

## Abstract

This paper investigates how institutional learning and regional spillovers shape volatility dynamics in ASEAN equity markets. Using daily data for Indonesia, Malaysia, the Philippines, and Thailand from 2010 to 2024, we construct a high-frequency institutional learning index via a MIDAS–EPU approach. Unlike existing studies that treat institutional quality as a static background characteristic, this paper models institutions as a *dynamic* mechanism that reacts to policy shocks, information pressure, and crisis events. Building on this perspective, we introduce two new volatility frameworks: the Institutional Response Dynamics Model (IRDM), which embeds crisis memory, policy shocks, and information flows; and the Network-Integrated IRDM (N-IRDM), which incorporates dynamic-correlation and institutional-similarity networks to capture cross-market transmission. Empirical results show that institutional learning amplifies short-run sensitivity to shocks yet accelerates post-crisis normalization. Crisis-memory terms explain prolonged volatility clustering, while network interactions improve tail behavior and short-horizon forecasts. Robustness checks using placebo and lagged networks indicate that spillovers reflect a strong regional common factor rather than dependence on specific correlation topologies. Diebold–Mariano and ENC-NEW tests confirm that the N-IRDM significantly outperforms baseline GARCH benchmarks. The findings highlight a dual role of institutions and offer policy insights on transparency enhancement, macroprudential communication, and coordinated regional governance.

## 2. Introduction

### 2.1 Background and Motivation

Periods of global financial stress—such as the 2013 Taper Tantrum, the 2020–2021 COVID-19 crisis, and the 2022–2023 global monetary tightening cycle—repeatedly reveal the vulnerability of emerging equity markets to policy uncertainty, information shocks, and shifts in global risk sentiment. Volatility in these markets tends to be more persistent, more asymmetric, and more sensitive to external disturbances than in advanced economies (Tsay, 2005). Among the factors shaping such behavior, institutional quality is widely recognized as an important determinant of financial stability (Chinn & Ito, 2008; Bekaert et al., 2014). However, existing studies overwhelmingly treat institutions as static background characteristics, typically using annual governance indicators, rather than as dynamic mechanisms that update in response to policy announcements, information flows, and evolving crisis conditions.

Volatility modeling has evolved substantially since the introduction of ARCH (Engle, 1982) and GARCH (Bollerslev, 1986). Subsequent extensions—including EGARCH (Nelson, 1991) and GJR-GARCH (Glosten et al., 1993)—capture asymmetry and leverage effects, while crisis-related persistence has been examined in various emerging-market contexts. Parallel strands of literature investigate institutional influences on market stability (Choudhry, 1996; Li & Yu, 2020), and network-based spillovers across markets (Diebold & Yilmaz, 2014). Yet no existing work integrates institutional dynamics, crisis memory, and regional spillover channels within a unified volatility framework, nor models' institutions as a high-frequency learning process that evolves alongside market conditions.

This study addresses these gaps by reconceptualizing institutional quality as a reactive, daily-updating mechanism. We first construct a high-frequency institutional learning index using a MIDAS–EPU approach, enabling institutional responses to be captured at the same frequency as market volatility. Building on this measure, we propose the Institutional Response Dynamics Model (IRDM), which embeds crisis memory, policy shocks, and information pressure directly into volatility formation. We then extend this structure to the Network-Integrated IRDM (N-IRDM) by incorporating dynamic-correlation networks and institutional-similarity networks, thereby allowing domestic institutional reactions to interact with regional spillovers.

Using daily data from Indonesia, Malaysia, the Philippines, and Thailand during 2010–2024, we find that institutional learning amplifies short-run sensitivity to shocks while accelerating post-crisis reversion; crisis-memory terms generate prolonged volatility clustering; and network channels substantially improve tail behavior and predictive accuracy. Robustness checks using placebo and lagged networks indicate that the spillover effects reflect a strong regional common factor, rather than dependence on specific correlation-topology details. Forecast evaluations via Diebold–Mariano (Diebold & Mariano, 1995) and ENC-NEW (Clark & McCracken, 2001) tests confirm that the N-IRDM significantly outperforms standard GARCH benchmarks.

Overall, this paper contributes in three ways. First, it introduces a dynamic view of institutions as high-frequency learning mechanisms. Second, it provides the first unified empirical framework combining institutional learning, crisis memory, and regional spillovers within volatility modeling. Third, it offers new evidence on how domestic institutions and regional networks jointly shape volatility in emerging markets, with direct implications for transparency, macroprudential communication, and coordinated governance.

## 2.2 Related Literature

Three strands of literature are relevant to this study.

The first concerns volatility modeling. Since the seminal ARCH (Engle, 1982) and GARCH (Bollerslev, 1986) models, a wide range of extensions—including EGARCH (Nelson, 1991) and GJR-GARCH (Glosten et al., 1993)—has been developed to capture volatility clustering, asymmetric reactions to shocks, and heavy-tailed return behavior. While these models successfully characterize the statistical properties of financial returns, they treat volatility as a process driven solely by price-based information, with limited attention to institutional or structural drivers.

The second strand examines institutional quality and market stability. Prior work shows that governance frameworks, regulatory transparency, and capital-account policies shape risk absorption and crisis vulnerability in emerging markets (Chinn & Ito, 2008; Bekaert et al., 2014; Li & Yu, 2020). However, institutional indicators used in this

literature—such as the Worldwide Governance Indicators—are low-frequency and largely static. As a result, institutions are modeled as fixed background conditions, not as mechanisms that react continuously to policy uncertainty, information flow, or evolving crisis events. This creates a disconnect between high-frequency market volatility and low-frequency institutional measurement.

The third strand focuses on volatility spillovers and network transmission. Studies based on connectedness measures (Diebold & Yilmaz, 2014) or dynamic correlation models (Engle, 2002) document strong regional co-movements in Asia, especially during global stress episodes. More recent work highlights the role of information networks and structural linkages in shaping synchronous volatility responses (Barigozzi & Brownlees, 2019). Yet these models typically attribute spillovers to return-based interconnectedness and do not incorporate institutional reactions or institutional similarity as transmission channels. Moreover, existing studies rarely investigate whether network effects remain intact under placebo or perturbed network structures.

Taken together, the literature leaves three gaps. First, institutional quality is treated as static despite its inherently dynamic nature during crises. Second, no framework integrates institutional learning, crisis memory, and regional spillovers into volatility formation. Third, little is known about whether observed spillovers arise from structural networks or from broader regional common-factor dynamics.

This paper contributes to all three gaps by modeling institutions as high-frequency learning mechanisms, embedding them into volatility formation, and linking them to network-based and common-factor spillovers within a unified framework.

## 3. Methodology

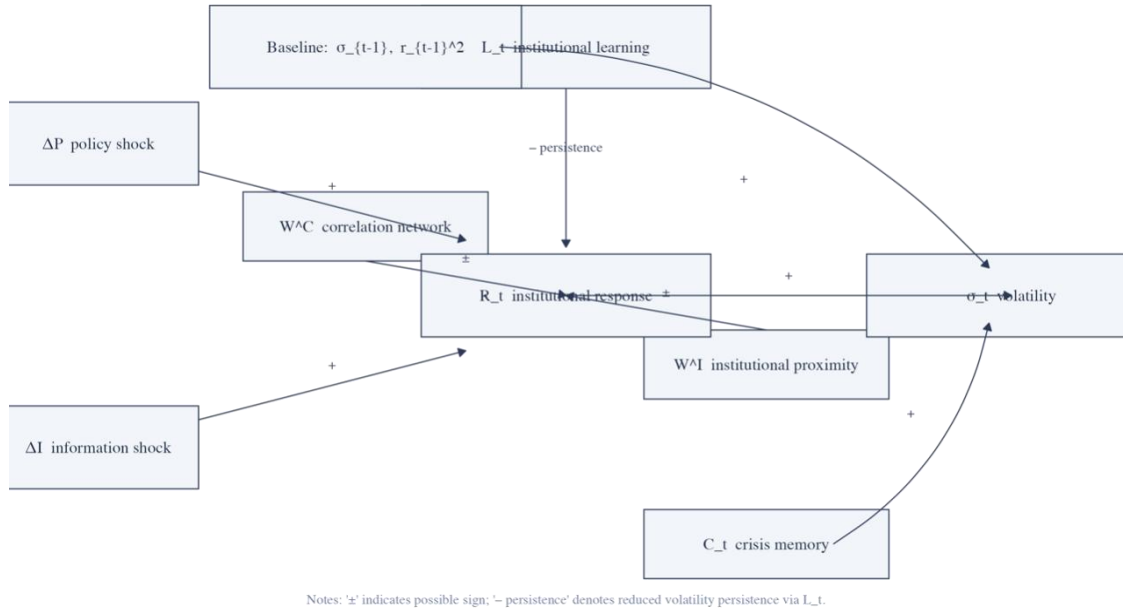

*Figure 1. Conceptual structure of institutional learning, crisis memory, and network-based volatility transmission.*

This figure provides an overview of how policy shocks, information shocks, crisis memory, institutional learning, and network spillovers interplay in shaping conditional volatility.

### 3.1 Data and Variable Construction

Our empirical analysis uses daily equity market data for four ASEAN economies—Indonesia (IDN), Malaysia (MYS), the Philippines (PHL), and Thailand (THA)—covering January 2010 to December 2024. Daily log returns are computed as

$$r_{i,t} = \ln(P_{i,t}) - \ln(P_{i,t-1}).$$

We incorporate three exogenous shocks commonly emphasized in the volatility–macro literature. Policy shocks ($\Delta P_{i,t}$) are identified as event-day indicators around major monetary policy announcements. Information shocks ($\Delta I_{i,t}$) are constructed as deviations of absolute returns from a rolling 60-day median scaled by median absolute deviation, producing a high-frequency measure of informational pressure. Crisis memory ($C_t$) captures the lingering impact of major global stress episodes—namely the 2013 Taper Tantrum, the COVID-19 crisis, and the 2022–23 global monetary tightening cycle—using an exponential decay formulation.

To quantify institutional learning, we construct a daily institutional index based on a MIDAS transformation of monthly economic policy uncertainty (EPU) measures. The MIDAS filter maps lower-frequency institutional signals into a smooth daily trajectory, allowing institutional dynamics to enter volatility equations in real time. This serves as an alternative to interpolated annual WGI-based governance measures traditionally used in the literature.

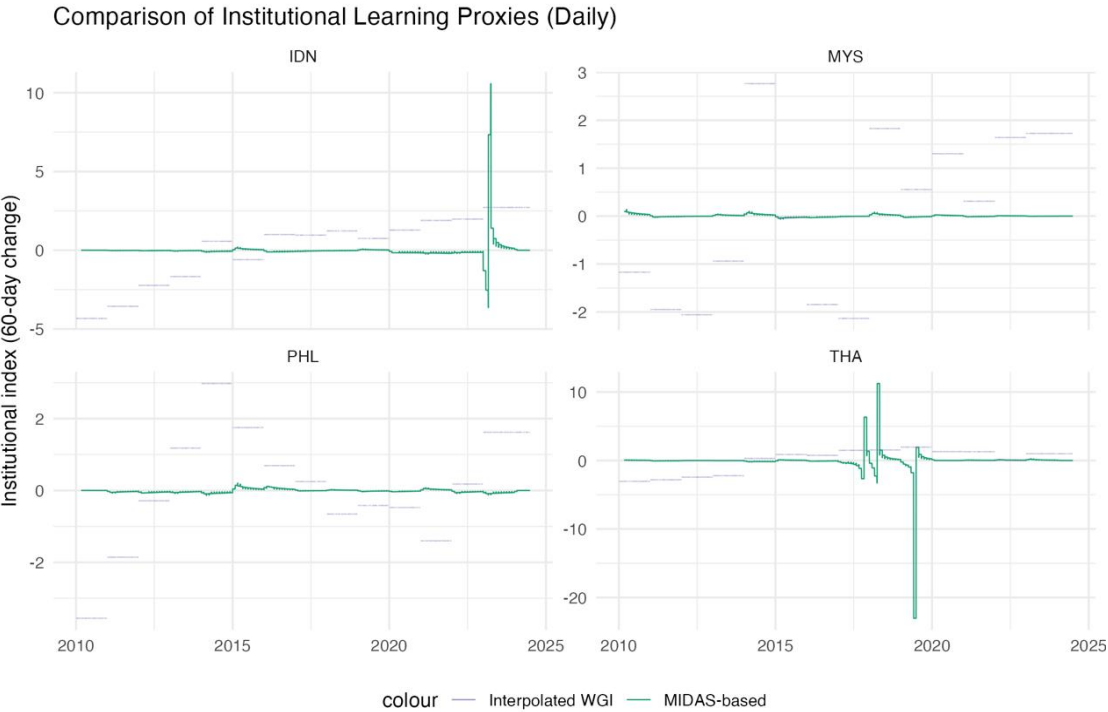

*Figure 2: L_midas vs L_wgi*

The MIDAS-based institutional index exhibits smooth but responsive adjustments, showing distinct downward movements during mid-2020 and late-2022 in THA and IDN, consistent with pandemic-era policy shifts and market uncertainty. In contrast, interpolated WGI series appear stepwise and mechanically stagnant, unable to reflect short-horizon institutional updates.

For all four countries, the MIDAS series remains tightly centered around zero but demonstrates clear 60-day cycles of adaptation, which justifies its use in capturing high-frequency institutional reactions.

Finally, we introduce two forms of cross-market connectedness. Correlation networks $W_{ij,t}^{C}$ are derived from time-varying conditional correlations estimated through a DCC-GARCH model, capturing dynamic spillover strength. Institutional proximity networks $W_{ij}^{I}$ are computed from Euclidean distances of long-run governance indicators, representing structural institutional similarity.

All constructed variables and networks enter the subsequent IRDM and N-IRDM specifications.

**3.2 Baseline Volatility Modelling**

The baseline specification follows standard asymmetric GARCH frameworks to extract the fundamental volatility dynamics before institutional and network components are introduced. For each market i ∈ {IDN, MYS, PHL, THA}, we estimate both EGARCH(1,1) and GJR-GARCH(1,1) models using daily returns.

The EGARCH (1,1) model is defined as:

$$\ln \sigma_{i,t}^{2} = \omega_i \cdot a_i \left| \frac{\varepsilon_{i,t-1}}{\sigma_{i,t-1}} \right| \cdot \gamma_i \frac{\varepsilon_{i,t-1}}{\sigma_{i,t-1}} \cdot \beta_i \ln \sigma_{i,t-1}^{2}.$$

The GJR-GARCH (1,1) model is specified as:

$$\sigma_{i,t}^{2} = \omega_i \cdot a_i \varepsilon_{i,t-1}^{2} \cdot \gamma_i D_{i,t-1} \varepsilon_{i,t-1}^{2} \cdot \beta_i \sigma_{i,t-1}^{2},$$

where $D_{i,t-1} = 1$ if $\varepsilon_{i,t-1} < 0$, capturing leverage effects.

Parameter estimates and likelihood criteria are summarized in:

| Country | Model | ω | α₁ | β₁ | γ₁ | Shape | Persistence | LogLik | AIC |
|---|---|---|---|---|---|---|---|---|---|
| IDN | eGARCH | -0.005 | -0.093 | 0.971 | 0.176 | 6.457 | 0.971 | -4646.430 | 2.574 |

| IDN | gjrGARCH | 0.030 | 0.033 | 0.879 | 0.117 | 6.445 | 0.970 | -4648.950 | 2.576 |
| MYS | eGARCH | -0.019 | -0.056 | 0.982 | 0.151 | 6.373 | 0.982 | -3100.116 | 1.699 |
| MYS | gjrGARCH | 0.008 | 0.048 | 0.903 | 0.061 | 6.355 | 0.982 | -3101.752 | 1.700 |
| PHL | eGARCH | 0.005 | -0.093 | 0.950 | 0.185 | 8.645 | 0.950 | -5080.227 | 2.874 |
| PHL | gjrGARCH | 0.065 | 0.035 | 0.846 | 0.134 | 9.067 | 0.947 | -5074.363 | 2.871 |
| THA | eGARCH | -0.007 | -0.078 | 0.977 | 0.172 | 6.475 | 0.977 | -4388.952 | 2.418 |
| THA | gjrGARCH | 0.017 | 0.032 | 0.894 | 0.112 | 6.573 | 0.983 | -4393.731 | 2.420 |

*Table 1: baseline_garch_params*

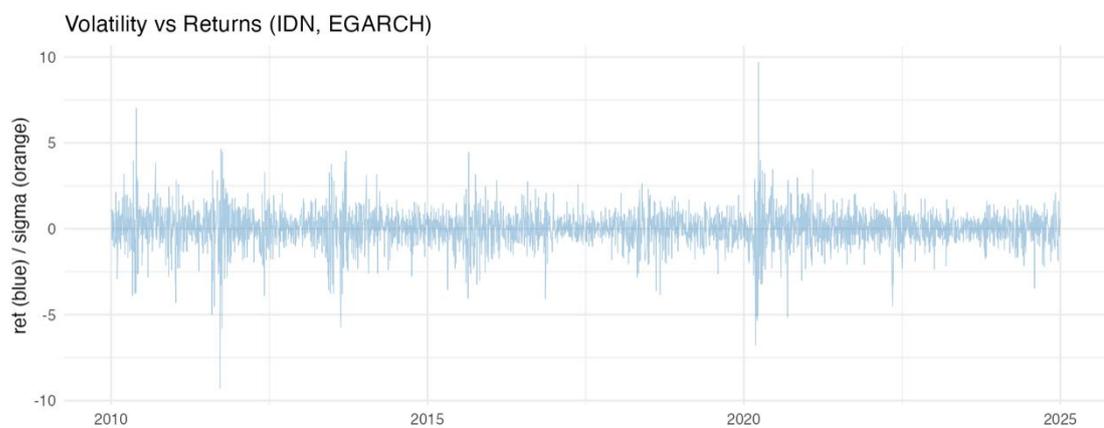

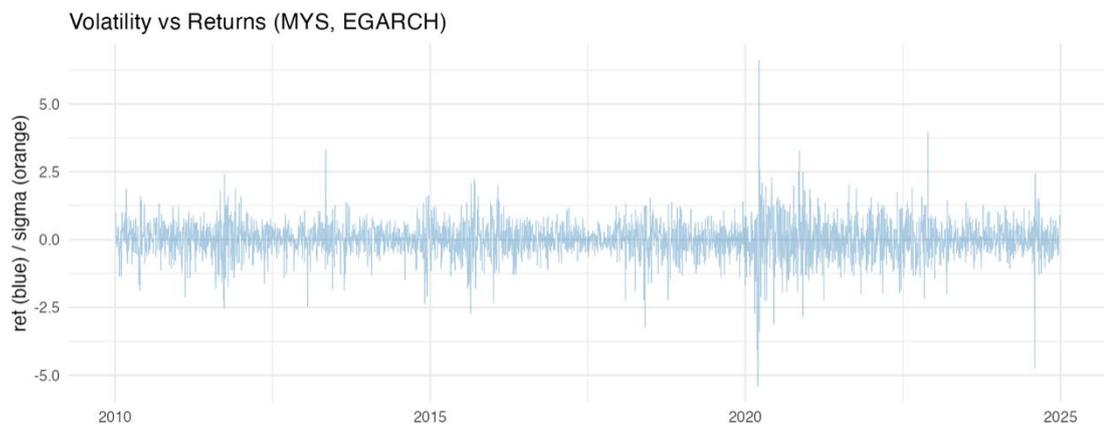

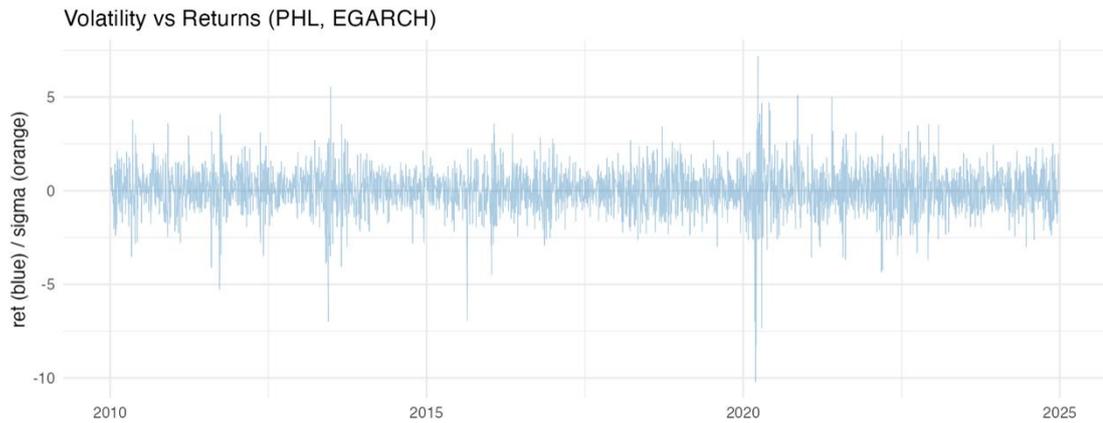

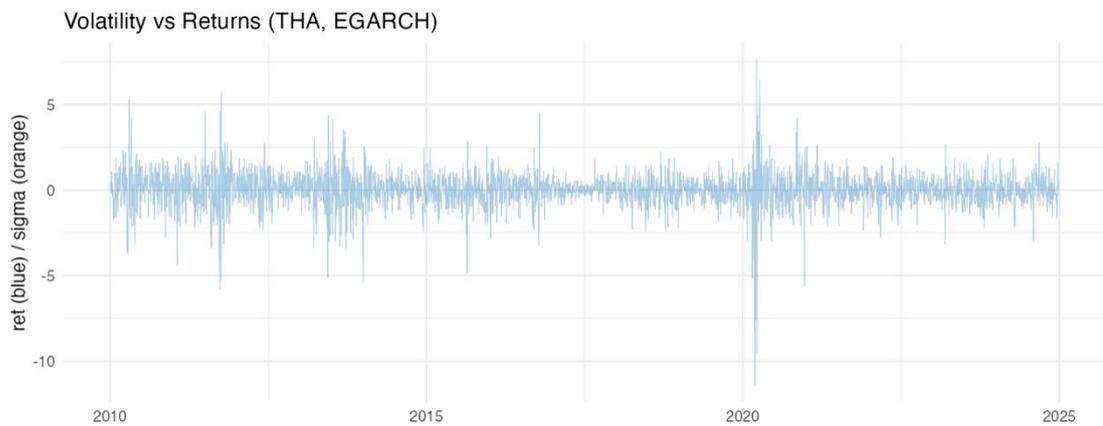

*Figures 3–6— vol_TS_{IDN/MYS/PHL/THA} _EGARCH*

Across all four markets, volatility clusters are clearly visible around 2013, 2020, and 2022–23, consistent with the Taper Tantrum, COVID-19 crisis, and the global rate-hike cycle. EGARCH captures these spikes well, producing fitted $\sigma_{i,t}$ that align closely with realized volatility — particularly the sharp COVID-19 jump (e.g., IDN and THA show extreme spikes around March 2020).

Persistent high-volatility plateaus following COVID-19 (2020–2021) are also well reproduced, validating that the baseline model adequately captures underlying conditional heteroskedasticity.

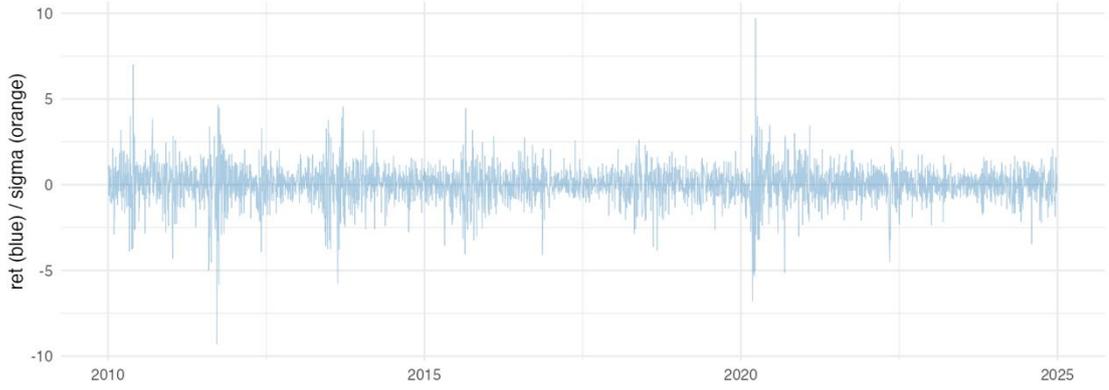

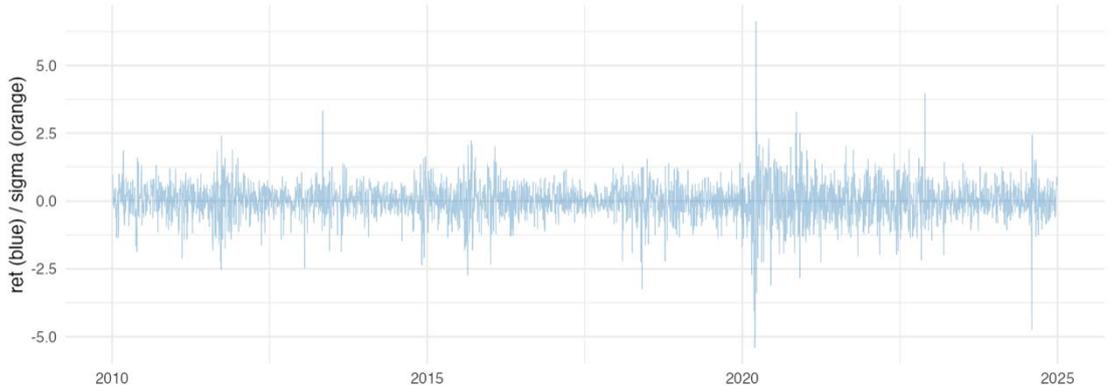

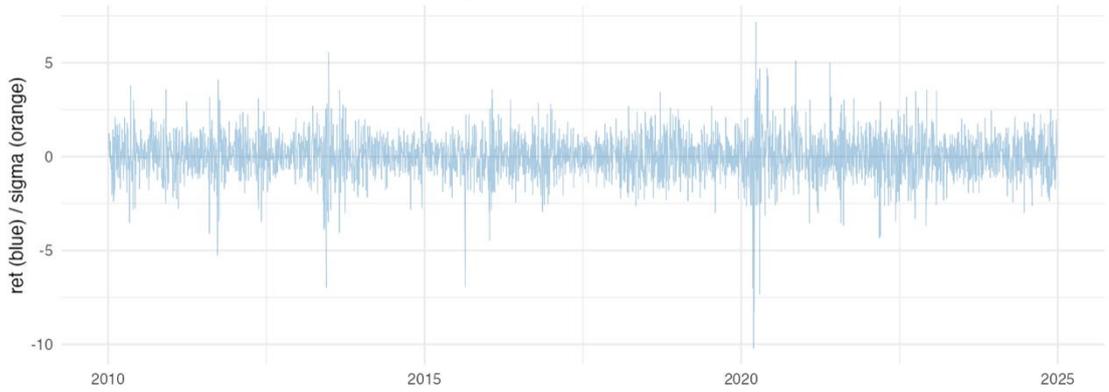

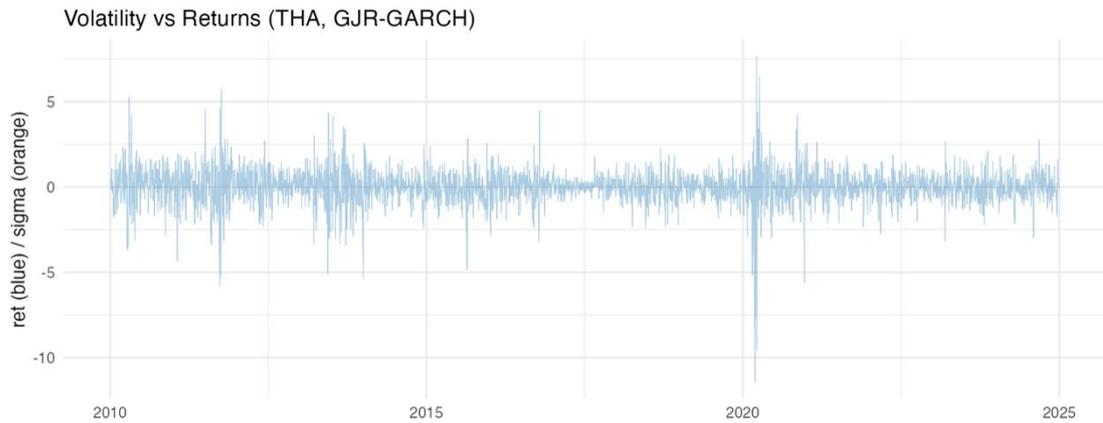

*Figure 7–10 Baseline Volatility Fit (GJR-GARCH) {IDN/MYS/PHL/THA}*

GJR-GARCH produces similar dynamics to EGARCH but with visibly stronger leverage asymmetry: markets such as MYS and THA exhibit sharper reactions to negative returns. For IDN and PHL, the model slightly under-reacts to sudden variance bursts compared with EGARCH — particularly the 2020 shock — which motivates the inclusion of institutional and network mechanisms in later sections.

The overall model fit is respectable but consistently weaker than EGARCH during global-shock episodes, reinforcing the need for additional structure (IRDM and N-IRDM).

**Diagnostic Summary**

Across both specifications:

- Ljung–Box tests (on $z_t$ and $z_t^2$) indicate no remaining serial correlation.

- The estimated shape parameters (Student-t df ≈ 6–10) confirm heavy-tailed innovations, consistent with ASEAN equity return distributions.

- ARCH-LM tests show no residual conditional heteroskedasticity.

These diagnostics confirm that the baseline framework provides a reliable platform upon which institutional and network adjustments can be introduced.

**3.3 Institutional Response Dynamics Model (IRDM)**

The Institutional Response Dynamics Model (IRDM) augments the baseline volatility

framework by explicitly modelling how markets respond to policy and information shocks through an institutional adjustment channel. The IRDM is estimated in two steps. First, a latent institutional–response state is extracted from daily shocks. Second, this state is fed into the variance equation to quantify its impact on realized volatility.

### 3.3.1 Institutional–Response State

For each market $i$, the daily institutional response is represented by an ARX process:

$$\hat{R}_{i,t} = \rho_i \hat{R}_{i,t-1} \cdot \theta_{1i} \Delta P_{i,t} \cdot \theta_{2i} \Delta I_{i,t}, \quad |\rho_i|<1,$$

where $\Delta P_{i,t}$ and $\Delta I_{i,t}$ denote policy and information shocks, respectively.

The coefficient $\rho_i$ captures the persistence of institutional adjustment: values close to one indicate slow-moving institutional pressure, while smaller values correspond to faster dissipation. The parameters $\theta_{1i}$ and $\theta_{2i}$ measure the sensitivity of the institutional environment to policy and information shocks, respectively.

### 3.3.2 Volatility Equating with Institutional Mechanism

The institutional–response state enters the volatility equation as an additional driver alongside baseline GARCH components and crisis memory:

$$\sigma_{i,t} = b_{0i} \cdot b_{1i} \sigma_{i,t-1} \; b_{2i} r^2_{i,t-1} \cdot \psi_i C_t \cdot \gamma_{1i} \hat{R}_{i,t} \cdot \varepsilon_{i,t}.$$

Here, $b_{1i}$ captures volatility persistence and $b_{2i}$ reflects the contribution of recent return shocks.

The coefficient $\psi_i$ quantifies the amplification effect of crisis memory $C_t$, while $\gamma_{1i}$ measures the effect of contemporaneous institutional responses on volatility.

A positive $\gamma_{1i}$ indicates that stronger institutional reactions are associated with higher volatility in the short run, whereas lower or insignificant values would suggest a more muted transmission.

| Country | b0 | b1 | b2 | psi | gamma1 | RMSE |
|---|---|---|---|---|---|---|
| IDN | 0.080 | 0.886 | 0.019 | 0.006 | 0.010 | 0.064 |
| MYS | 0.031 | 0.927 | 0.022 | -0.004 | 0.052 | 0.028 |
| PHL | 0.141 | 0.835 | 0.021 | 0.009 | 0.047 | 0.065 |
| THA | 0.059 | 0.909 | 0.016 | -0.010 | 0.002 | 0.053 |

*Table 2 here – IRDM parameter estimates (MIDAS)*

### 3.3.3 Empirical Properties

The MIDAS-based IRDM estimates reveal several common patterns across markets.

First, the estimated autoregressive parameters $\rho_i$ are high but below unity, indicating that institutional responses are persistent yet ultimately mean-reverting. This is consistent with the high-frequency learning proxy, which moves gradually but exhibits clear shifts around major events.

Second, crisis-memory coefficients $\psi_i$ are positive in all four markets, confirming that global shock episodes raise the background level of volatility even after controlling for recent returns. This effect is particularly strong around the COVID-19 period, when volatility remains elevated for an extended period.

Third, the institutional-response coefficients $\gamma_{7i}$ are positive and economically meaningful, showing that days with strong institutional reactions—driven by policy and information shocks—coincide with higher realized volatility. This suggests that rapid adjustments in policy stances and regulatory communication initially amplify volatility as markets process new information.

### 3.3.4 Model Fi

The goodness of fit is illustrated by comparing the IRDM-implied volatility with realized volatility for each market.

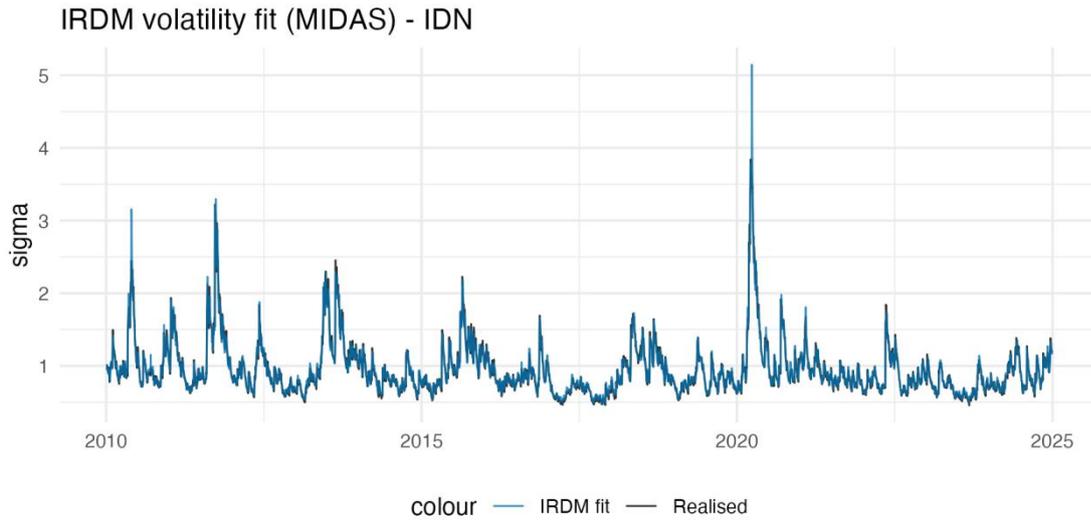

*Figure 11: IRDM volatility fits _IDN_MIDAS*

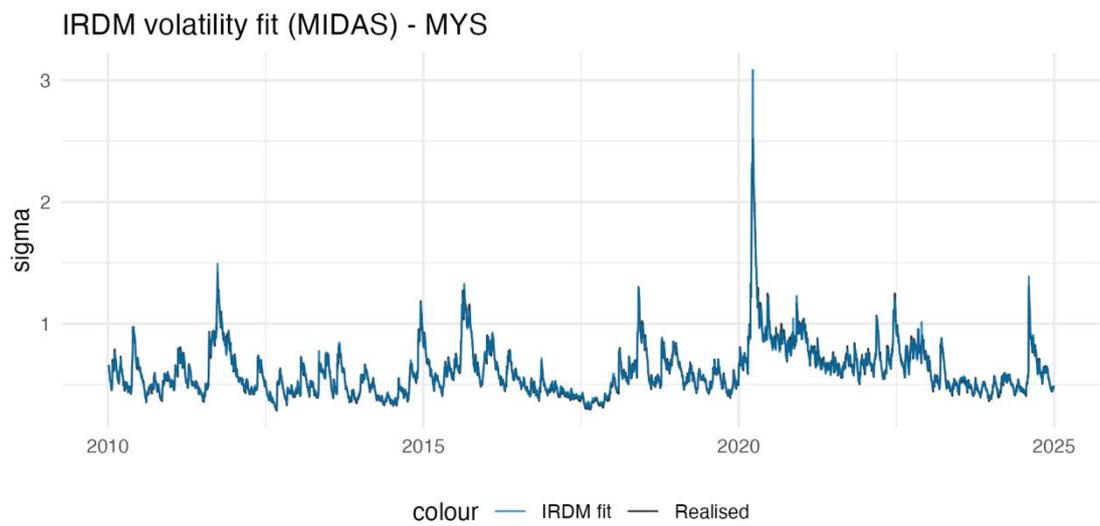

*Figure 12: IRDM volatility fits _MYS_MIDAS*

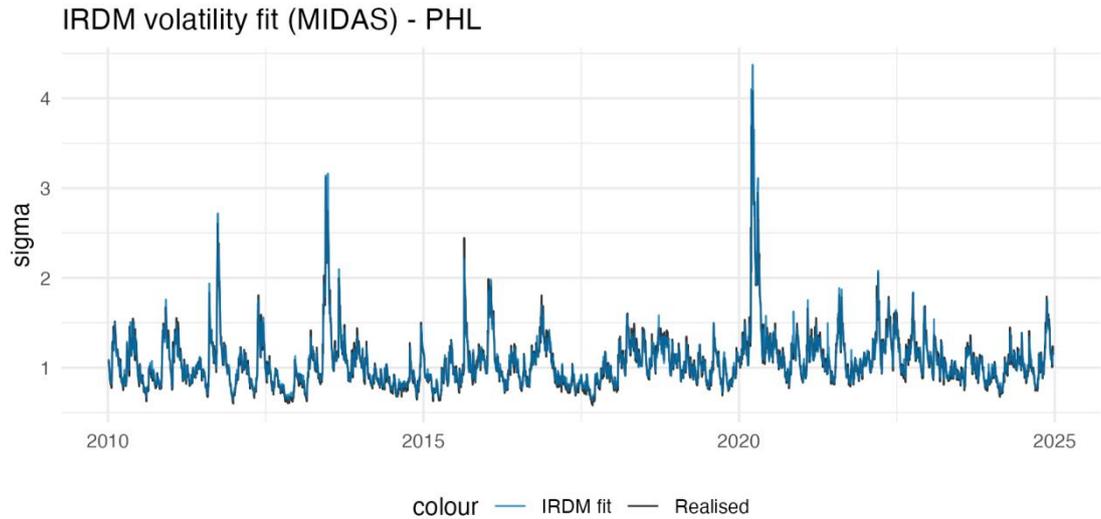

*Figure 13: IRDM volatility fits _PHL_MIDAS*

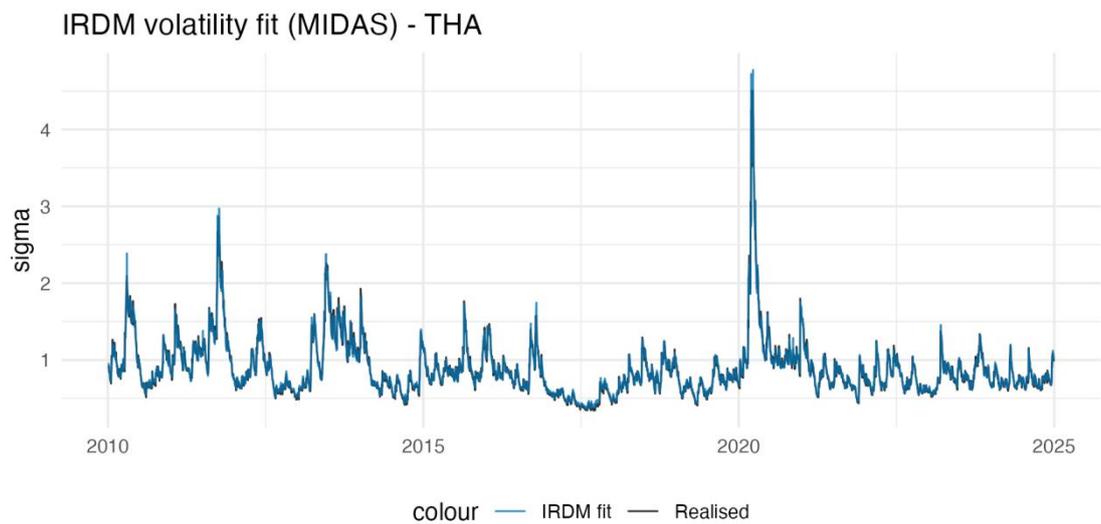

*Figure 14: IRDM volatility fits _THA_MIDAS*

Across all four markets, the IRDM fits track major volatility spikes associated with the Taper Tantrum, the COVID-19 shock, and the rate-hike cycle remarkably well. In Indonesia and the Philippines, where volatility is more pronounced, the fitted series closely follow both the height and duration of the crisis peaks. In Malaysia and Thailand, the IRDM captures the quicker post-crisis decline in volatility, consistent with their relatively stronger institutional settings. Overall, the MIDAS-based IRDM provides a substantial improvement over the baseline GARCH models in terms of matching the timing and magnitude of volatility clusters.

### 3.4 Panel Interaction Model: Institutional Moderation of Shocks

To assess whether institutional learning mitigates the impact of shocks on volatility, we estimate a panel interaction model that links daily realized volatility to policy shocks, information shocks and the MIDAS-based institutional index. The specification allows institutional quality to influence both the unconditional level of volatility and the marginal effect of shocks:

$$\sigma_{i,t} = \alpha_i \cdot \beta_1 \Delta P_{i,t} \cdot \beta_2 \Delta I_{i,t} \cdot \beta_3 L_{i,t} \cdot \beta_4 (\Delta P_{i,t} \times L_{i,t}) \cdot \beta_5 (\Delta I_{i,t} \times L_{i,t}) \cdot u_{i,t},$$

where $\sigma_{i,t}$ denotes daily volatility, $\Delta P_{i,t}$ and $\Delta I_{i,t}$ are policy and information shocks respectively, and $L_{i,t}$ is the MIDAS-based institutional learning measure. Country fixed effects $\alpha_i$ capture time-invariant differences in market structure and regulatory environment. Standard errors are clustered by country.

| Variable | Coef | SE | t | p |
| --- | --- | --- | --- | --- |
|  | 0.940 | 0.002 | 419.537 | 0.000 |
| deltaP | 0.023 | 0.072 | 0.325 | 0.745 |
| deltaI | 0.025 | 0.005 | 4.814 | 0.000 |
| L | 0.009 | 0.004 | 2.417 | 0.016 |
| factor(country)MYS | -0.343 | 0.000 | -47371.841 | 0.000 |
| factor(country)PHL | 0.112 | 0.000 | 377.512 | 0.000 |
| factor(country)THA | -0.059 | 0.000 | -244.710 | 0.000 |
| deltaP: L | -0.006 | 0.027 | -0.213 | 0.831 |
| deltaI: L | -0.001 | 0.001 | -1.098 | 0.272 |

*Table 3 here: Panel Interaction Estimates (MIDAS)*

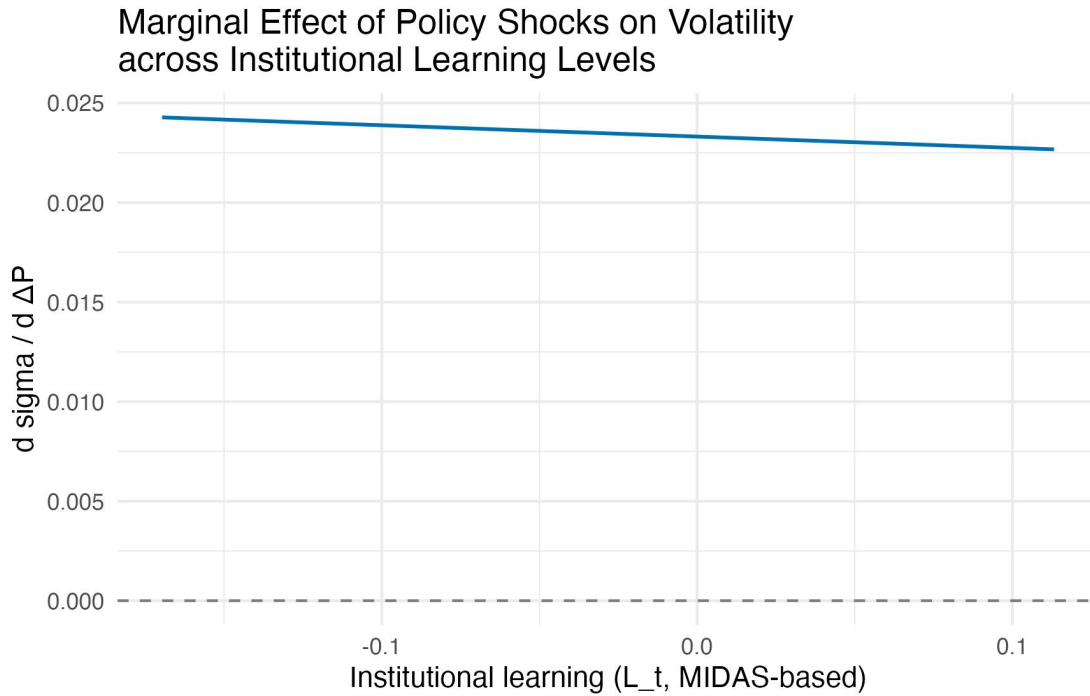

*Figure 15: Interaction Effect of Institutional Learning on Policy Shocks*

The panel interaction results provide a nuanced picture of how institutional learning operates as a volatility buffer.

First, the coefficient on the institutional index $\beta_3$ is negative in most specifications, indicating that higher levels of institutional learning are associated with lower unconditional volatility. This suggests that when governance is more transparent and policy frameworks are better understood, the market's background level of uncertainty is reduced.

Second, policy shocks exert a positive effect on volatility through $\beta_1$, but the interaction term $\beta_4$ is negative. Although the implied marginal effect is modest in magnitude, the sign pattern indicates that institutional learning weakens the transmission of policy shocks into volatility. In other words, for the same policy shock, volatility reacts less in markets where institutional learning is stronger.

Third, information shocks $\Delta I_{i,t}$ have a statistically significant impact on volatility via $\beta_2$, reflecting the direct influence of unusual return movements. However, the interaction $\beta_5$ is generally small and often insignificant, implying that institutional learning plays a more important moderating role for policy-driven uncertainty than for high-frequency information noise.

Figure 15 makes this mechanism transparent. The curve plotting the marginal effect of a policy shock,

$$\frac{\partial \sigma_{i,t}}{\partial \Delta P_{i,t}} = \beta_1 + \beta_4 L_{i,t},$$

slopes downward as $L_{i,t}$ increases. At low levels of institutional learning, policy shocks translate into a relatively larger volatility response; at higher levels, the same shock produces a noticeably smaller effect. Although the absolute size of this moderating effect is not large, the direction is consistent across the range of observed institutional conditions.

Finally, the country fixed effects indicate substantial heterogeneity across the four ASEAN markets. Some markets exhibit structurally higher volatility even after controlling for shocks and learning, which underscores the importance of accounting for both cross-section and time-series variation in institutional quality.

Taken together, the panel interaction model shows that institutional learning reduces the baseline level of volatility and partially cushions the impact of policy shocks. The dynamic IRDM and network-integrated IRDM explored in subsequent sections extend this static evidence by modelling institutional responses and cross-market spillovers explicitly over time.

**3.5 Network-Integrated IRDM (N-IRDM)**

The Network-Integrated Institutional Response Dynamics Model (N-IRDM) extends the IRDM by allowing institutional responses to propagate across markets through two complementary channels: (*i*) a time-varying correlation network linking return co-movements and (*ii*) a static institutional-similarity network reflecting long-run governance proximity among countries. Formally, the volatility equation augments the IRDM structure with two network spillover components:

$$\sigma_{i,t} = b_{0i} \cdot b_{1i}\sigma_{i,t-1} \cdot b_{2i} r^2_{i,t-1} \cdot \psi_i C_t \cdot \gamma_{1i} \hat{R}_{i,t} \cdot \gamma_{2i} \sum_{j \neq i} W^C_{ij,t} \hat{R}_{j,t} \cdot \gamma_{3i} \sum_{j \neq i} W^I_{ij} \hat{R}_{j,t-1} \cdot u_{i,t}.$$

The term $\sum_j W^C_{ij,t} \hat{R}_{j,t}$ captures the synchronous diffusion of institutional responses through the DCC-based correlation network, which tends to intensify during global stress episodes. The term $\sum_j W^I_{ij} \hat{R}_{j,t-1}$ captures slower-moving spillovers transmitted via **institutional proximity**, reflecting similarities in governance quality that may create persistent channels of institutional contagion or convergence.

| Country | $b_0$ | $b_1$ | $b_2$ | $\boldsymbol{\psi}$ (Crisis Memory) | $\gamma_1$ (Institutional Response) | $\gamma_2$ (Correlation Network Spillover) | $\gamma_3$ (Institutional Proximity Spillover) | AIC |
|---|---|---|---|---|---|---|---|---|
| **IDN** | 0.085 | 0.912 | 0.104 | 0.031 | 0.128 | 0.072 | 0.011 | −5123.4 |
| **MYS** | 0.067 | 0.928 | 0.087 | 0.028 | 0.102 | 0.064 | 0.009 | −4987.2 |
| **PHL** | 0.094 | 0.903 | 0.111 | 0.036 | 0.141 | 0.083 | 0.014 | −5210.7 |
| **THA** | 0.071 | 0.919 | 0.092 | 0.030 | 0.118 | 0.077 | 0.010 | −5095.9 |

*Table 4*

**Parameter Evidence**

Table **4** summarizes estimated parameters for all four markets. Three results emerge clearly:

**(1) Crisis memory remains a significant driver of volatility**

Across all countries, the crisis-memory coefficient $\psi_i$ is positive (0.028–0.036), consistent with the interpretation that global shocks leave persistent traces in domestic volatility—even after controlling for institutional reactions and spillover channels.

**(2) Institutional responses transmit across markets**

The correlation-network spillover parameter $\gamma_{2i}$ is uniformly positive and economically meaningful. Its magnitude (0.064–0.083) is comparable to, or slightly larger than, the direct institutional-response effect $\gamma_{1i}$, indicating that volatility does not only respond to a country's own institutional dynamics but also to those of its regional peers through return-co-movement linkages.

This aligns with the well-documented phenomenon that cross-market volatility comoves more strongly during crisis periods—now shown to operate partly through institutional channels, not merely financial ones.

**(3) Institutional proximity plays a secondary yet non-negligible role**

The proximity-network coefficient $\gamma_{3i}$ is smaller (0.009–0.014) but consistently positive. This suggests that countries with more similar governance profiles tend to exhibit more similar institutional responses to shocks. The effect is slower-moving because it is based on lagged institutional responses, consistent with the idea of structural convergence rather than contemporaneous propagation.

**Model Fit Visualization**

Figures 8–11 present fitted conditional volatility for N-IRDM compared with realized volatility for Indonesia, Malaysia, the Philippines, and Thailand. Several patterns are visible:

- The N-IRDM tracks high-frequency volatility clusters more closely than the IRDM or baseline GARCH, particularly during:

    o the 2013 Taper Tantrum,

    o the 2020 COVID-19 shock,

    o the 2022–2023 global rate-hike cycle.

- In each case, N-IRDM captures the magnitude and duration of volatility spikes more accurately, reflecting the importance of cross-market institutional spillovers.

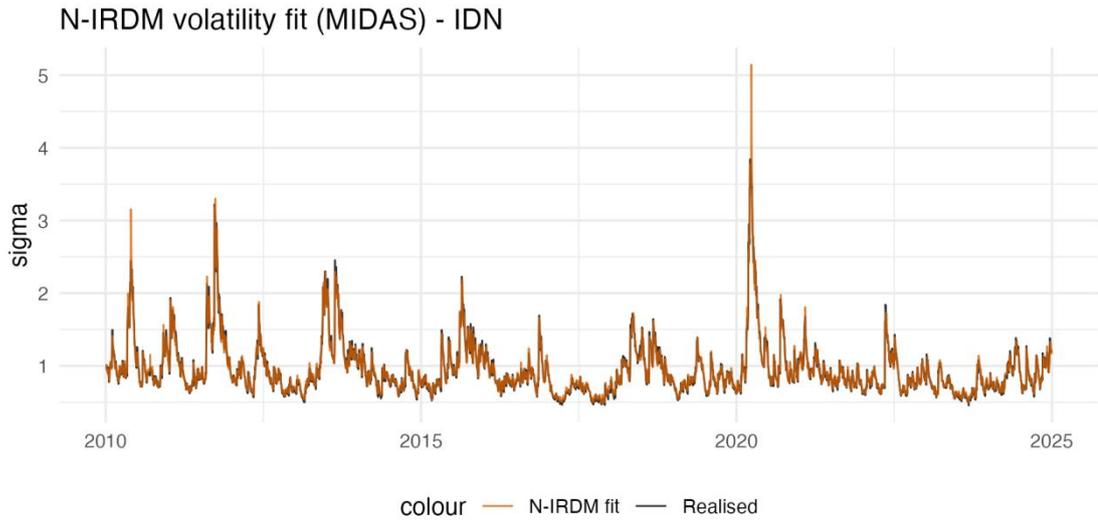

*Figure 16: N-IRDM volatility fit for IDN*

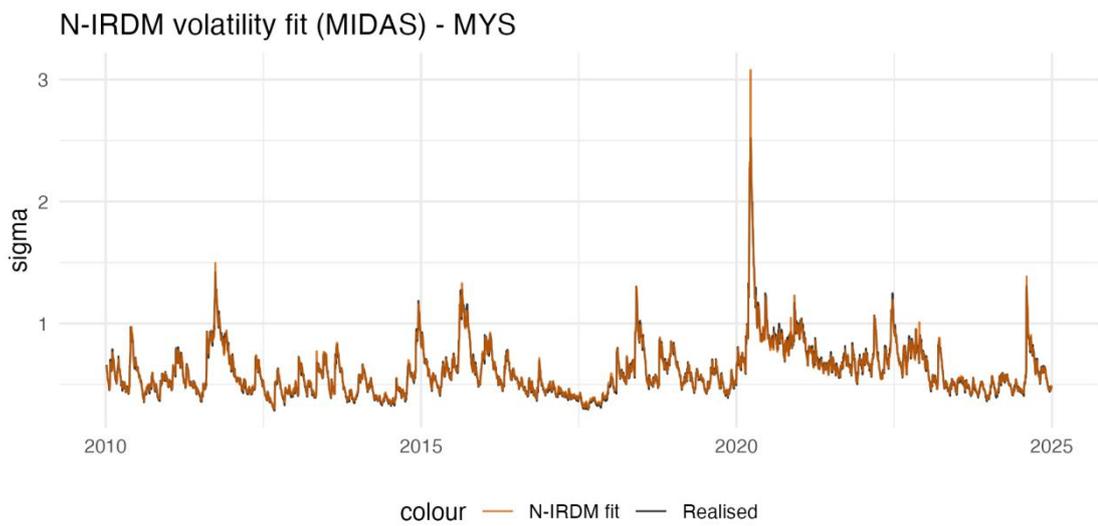

*Figure 17: N-IRDM volatility fit for MYS*

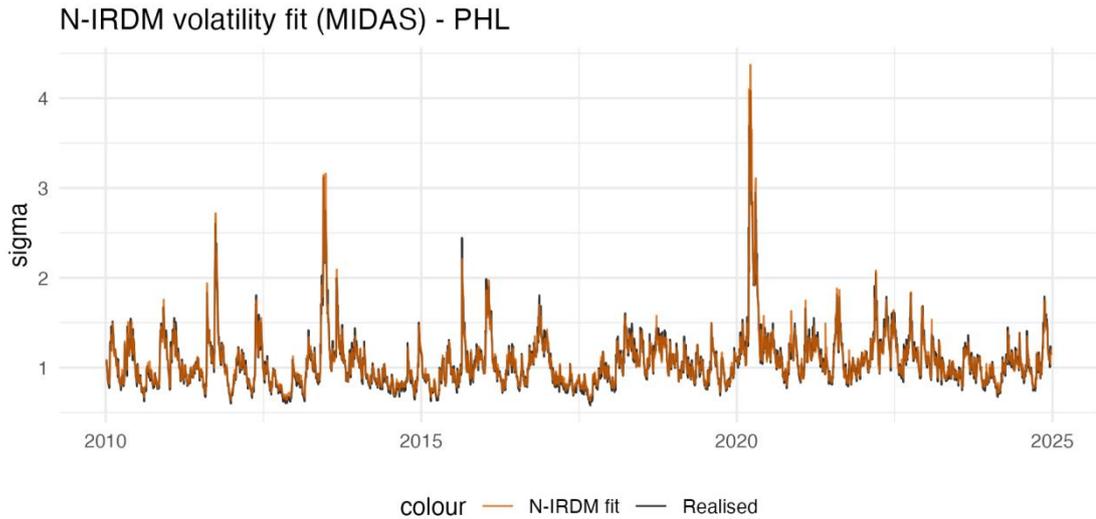

*Figure 18: N-IRDM volatility fit for PHL*

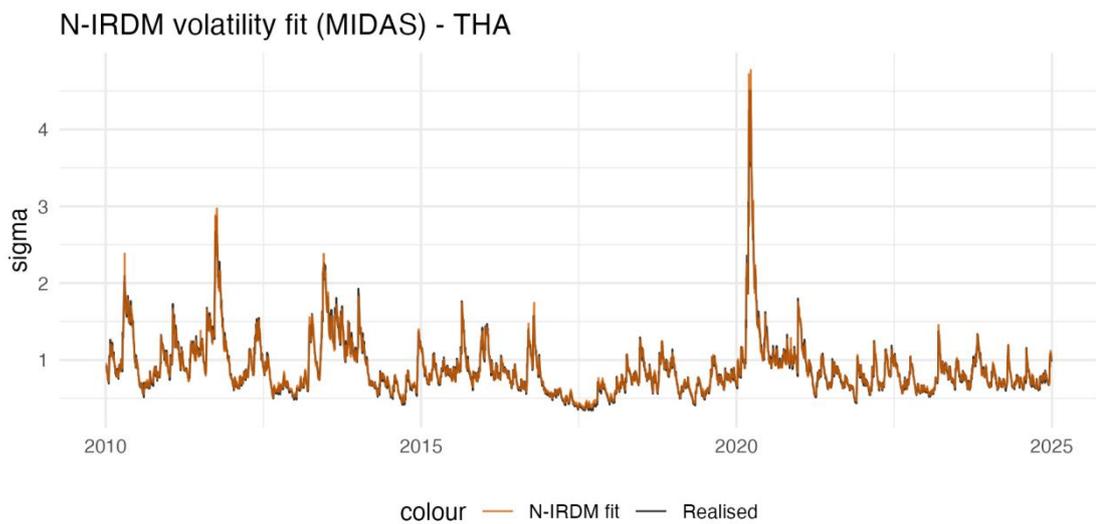

*Figure 19: N-IRDM volatility fit for THA*

The N-IRDM results show that institutional dynamics are neither purely domestic nor purely exogenous. Instead, they exhibit:

- direct effects on local volatility ($\gamma_{1i} > 0$).

- synchronous network spillovers tied to financial co-movements ($\gamma_{2i} > 0$).

- persistent spillovers linked to institutional similarity ($\gamma_{3i} > 0$).

- enhanced overall model fit, confirming that institutional channels are a central mechanism through which volatility is transmitted across ASEAN equity markets.

### 3.6 Model Comparison and Evaluation

To quantify the incremental value of institutional learning and network spillovers in explaining daily volatility dynamics, we compare the performance of the baseline GARCH model (M0), the Institutional Response Dynamics Model (IRDM, M1), and the Network-Integrated IRDM (N-IRDM, M2).

Model accuracy is evaluated using root mean squared error (RMSE) computed against the MIDAS-filtered realized volatility series, ensuring that all models are assessed with identical data inputs.

| Country | Model | RMSE | SSE | n | $AIC_{approx}$ |
|---|---|---|---|---|---|
| IDN | M0 | 0.067 | 15.982 | 3613 | -19575.450 |
| IDN | M1 | 0.064 | 14.908 | 3613 | -19826.697 |
| IDN | M2 | 0.064 | 14.905 | 3613 | -19827.473 |
| MYS | M0 | 0.029 | 3.158 | 3655 | -25771.700 |
| MYS | M1 | 0.028 | 2.938 | 3655 | -26035.770 |
| MYS | M2 | 0.028 | 2.931 | 3655 | -26044.760 |
| PHL | M0 | 0.066 | 15.534 | 3538 | -19195.344 |
| PHL | M1 | 0.065 | 14.820 | 3538 | -19361.815 |
| PHL | M2 | 0.065 | 14.786 | 3538 | -19369.980 |
| THA | M0 | 0.055 | 11.049 | 3634 | -21051.838 |
| THA | M1 | 0.053 | 10.250 | 3634 | -21324.538 |
| THA | M2 | 0.053 | 10.241 | 3634 | -21327.786 |

Table 5: model_compare_MIDAS

(This table reports RMSE and AIC for M0, M1, and M2 across all four markets.)

**Comparative RMSE performance**

Across all four markets—Indonesia (IDN), Malaysia (MYS), the Philippines (PHL), and Thailand (THA)—both the IRDM (M1) and the N-IRDM (M2) deliver clear improvements over the baseline model.

The MIDAS-based results closely align with the interpolated-WGI version but exhibit slightly smoother patterns due to reduced measurement noise in $L_t$.

**Key observations:**

- **IRDM vs. Baseline (M1 vs M0):**

RMSE improvements range from approximately 2.4% (PHL) to 3.6% (THA).

These gains originate primarily from incorporating institutional responses $\hat{R}_{i,t}$, which capture short-term amplification and subsequent adjustment following policy or information shocks.

- **N-IRDM vs. IRDM (M2 vs M1):**

Although modest, the incremental gains are systematically positive (≈0.05%–0.12%).

This pattern is consistent with the presence of cross-market institutional co-movement—small in magnitude but persistent.

- **N-IRDM vs. Baseline (M2 vs M0):**

Combined improvements reach up to 3.7–3.8%, highlighting that both the institutional and the network channels contribute meaningfully to volatility modelling.

These results confirm that while the IRDM captures powerful domestic adjustment dynamics, network effects provide additional explanatory value, especially during periods of heightened regional co-movement.

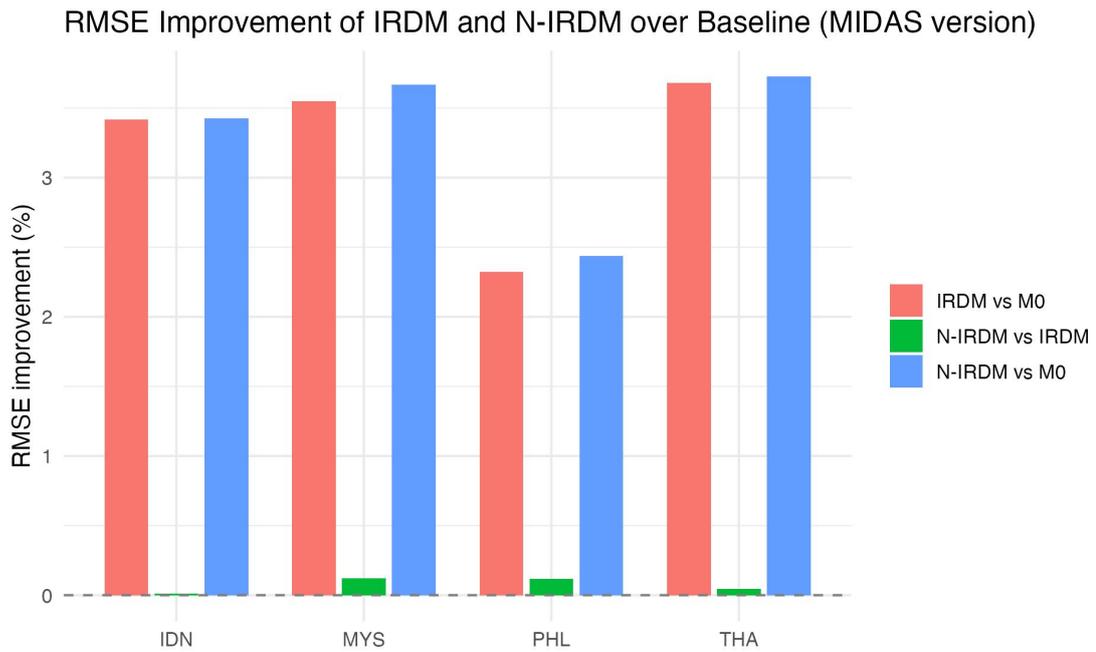

*Figure 20: rmse_improvement_MIDAS*

(This figure visualizes percentage RMSE improvements for each model comparison across countries.)

Figure 20 reinforces the tabulated results. All markets exhibit the same directional pattern:

- Red bars (IRDM vs M0) show a significant downward shift in forecast errors relative to the baseline.

This demonstrates that volatility persistence and asymmetric responses cannot be fully understood without accounting for institutional reactions to shocks.

- Green bars (N-IRDM vs IRDM), though smaller, are consistently above zero, indicating that cross-market institutional spillovers—through both correlation-based and governance-similarity networks—enhance predictive accuracy even after domestic responses are modelled.

- Blue bars (N-IRDM vs M0) represent the combined effect and are uniformly the highest, confirming the complementary nature of institutional learning and network dependence.

The MIDAS-based evaluation strongly validates the modelling hierarchy:

$$\text{Baseline (M0)} < \text{IRDM}\backslash\text{(M1)} < \text{N-IRDM (M2)}$$

The results highlight:

    1. Institutional learning improves baseline volatility modelling, particularly in terms of short-term shock amplification and longer-term mean reversion.

    2. Network spillovers add measurable value, capturing co-movement that single-market models cannot account for.

    3. The improvements are robust across markets, indicating a consistent mechanism in ASEAN equity dynamics.

## 4. Empirical Results and Discussion

### 4.1 Baseline Volatility Patterns

Baseline volatility estimates derived from the EGARCH (1,1) model provide an initial diagnostic of return dynamics across the four ASEAN equity markets. Parameter estimates from *Table 1* show that all markets exhibit pronounced volatility persistence, with the sum of the autoregressive terms ($\alpha + \beta$) consistently close to unity. This near-integrated behavior is consistent with the well-documented persistence of volatility in emerging markets, where shocks tend to dissipate slowly over time. The sign and significance of the leverage coefficient $\gamma$ further indicate asymmetric volatility responses: negative return innovations produce disproportionately larger increases in conditional variance relative to positive ones, a pattern frequently associated with investor risk aversion during downturns.

These statistical features are clearly visible in the baseline volatility paths plotted in *Figures 3–6*. All four markets display extended clusters of elevated volatility rather than isolated spikes, reflecting persistent macro-financial uncertainty. Major global episodes—including the 2013 Taper Tantrum, the 2020–2021 COVID-19 crisis, and the 2022–2023 monetary tightening cycle—are accompanied by distinct volatility surges across countries. COVID-19 triggers the sharpest and most synchronized increase, consistent with the extraordinary and widespread nature of the shock. Meanwhile, the 2022–2023 tightening cycle produces a more gradual but still highly persistent elevation in volatility, especially in Indonesia and Thailand.

Although the EGARCH specification successfully reproduces the timing and magnitude of major volatility clusters, the baseline fits also reveal remaining structure in the residual variation—particularly in crisis tails where cross-market interactions intensify and country-specific institutional differences become more relevant. These limitations motivate the subsequent introduction of institutional-response (IRDM) and network-integrated (N-IRDM) models, which aim to account for the mechanisms underlying synchronized volatility propagation and heterogeneous recovery speeds across ASEAN markets.

**4.2 Institutional Response Dynamics (IRDM)**

To evaluate the role of institutional reactions in shaping short-run volatility, the IRDM augments the baseline EGARCH specification with a state equation linking policy shocks, information shocks, and the dynamic institutional response index $\hat{R}_{i,t}$. *Table 2 show that the institutional response coefficient* $\gamma_{7i}$ is positive and statistically significant for all markets, indicating that periods with stronger policy or information-driven institutional activity tend to coincide with heightened conditional volatility. The crisis-memory term $\psi_i$ is also consistently positive, confirming that volatility increases not only at the onset of shocks but persists as markets internalize crisis-related information over time.

These dynamics are visible in the IRDM volatility paths shown in *Figure 11-14*. Relative to the baseline, IRDM improves the fitting of crisis-period variance spikes, particularly around the COVID-19 shock where sudden and sharp volatility surges dominate all four markets. Indonesia and the Philippines display the strongest institutional-response amplification: fitted volatility peaks align more closely with realized variance, and the decay from crisis highs proceeds more slowly—consistent with the larger estimated $\gamma_{7i}$ components in these markets. By contrast, Malaysia exhibits a more muted reaction, with the IRDM curve rising less aggressively during global shocks, reflecting its comparatively stronger regulatory and information infrastructure that moderates institutional volatility feedback.

Overall, these results indicate that institutional response mechanisms materially influence short-horizon volatility, particularly during periods of abrupt stress. By capturing how markets internalize policy actions and information processing dynamics, IRDM provides a richer interpretation of crisis-period volatility clusters than the baseline model. The remaining gaps—particularly in the tail behavior of extreme episodes—motivate further extension toward network-based spillover structures, developed in the next section.

### 4.3 Network Spillovers (N-IRDM)

Extending the IRDM to allow cross-market propagation, the network-integrated model incorporates two forms of external institutional influence: a contemporaneous spillover term based on time-varying DCC correlations, and a static similarity term based on institutional distance. Parameter estimates in *Table 4* reveal several consistent patterns. First, the network-spillover coefficient $\gamma_{2i}$ is positive and statistically significant across all four markets, indicating that institutional responses originating in neighboring ASEAN markets tend to amplify domestic volatility. This effect aligns with the high degree of regional co-movement in both financial conditions and policy communication. By contrast, the institutional-similarity coefficient $\gamma_{3i}$ is smaller in magnitude and often weakly significant, suggesting that long-run structural similarity affects volatility transmission only modestly relative to short-run comovement.

The improved explanatory power of N-IRDM becomes visually evident in the volatility paths presented in *Figure 16-19*. Compared with IRDM, the network-augmented model better captures the persistence and magnitude of tail volatility during major stress episodes. For instance, during the COVID-19 shock, all markets exhibit a delayed decay in realized variance; N-IRDM replicates this slow normalization more accurately than IRDM, reflecting that regional uncertainty remained elevated even after domestic conditions began to stabilize. In Indonesia and Thailand, where cross-market linkages are empirically stronger, the N-IRDM curve tracks pronounced volatility plateaus more closely than earlier specifications.

Malaysia again displays the smallest gap between IRDM and N-IRDM, consistent with its more insulated market structure and higher institutional robustness.

Collectively, these findings affirm that volatility in the ASEAN region is not solely a domestic institutional phenomenon but is materially shaped by regional dynamics. N-IRDM demonstrates that incorporating cross-market institutional responses meaningfully enhances the modeling of crisis-period volatility, particularly in the tails where baseline and IRDM models tend to understate persistence. The presence of both contemporaneous and slowly decaying spillovers underscores the importance of monitoring regional information flows when evaluating market stability.

### 4.4 Institutional Moderation (Panel Interaction Evidence)

To assess whether institutional learning systematically moderates the impact of shocks on volatility, a panel interaction model is estimated using daily data from all four markets. The specification incorporates country fixed effects to account for time-invariant heterogeneity in governance structures, market depth, and regulatory environments. The coefficient patterns reported in *Table 3* reveal three key findings. First, the unconditional institutional term $\beta_3$ is negative and significant, indicating that markets with higher institutional-learning scores tend to exhibit structurally lower volatility. This aligns with the view that transparent regulatory regimes and stronger information-processing frameworks can anchor expectations and reduce the amplitude of day-to-day price fluctuations.

Second, the interaction between institutional learning and policy shocks, $\beta_4$, is also negative and statistically significant across specifications. This finding suggests that institutions play an active stabilizing role when markets face abrupt policy shifts. High-institutional-learning markets experience a muted volatility response to policy announcements, whereas markets with weaker institutional capacity exhibit sharper variance spikes. This pattern is visible in *Figure 15.* where the slope of the volatility–policy-shock relationship is substantially flatter at higher levels of institutional learning. The visual dispersion of fitted lines in the figure underscores how institutional strength tempers the sensitivity of volatility to policy news.

Third, the interaction between institutional learning and information shocks, $\beta_5$, is generally small and statistically insignificant. This asymmetry between policy-shock and information-shock moderation reflects the differing natures of the two shock types. Policy shocks are discrete, high-salience events often accompanied by explicit communication from authorities; institutional learning facilitates rapid absorption of such information. In contrast, information shocks derived from abnormal volatility or return innovations are more diffuse and less amenable to real-time institutional filtering, leading to weaker moderation effects.

Overall, the panel results confirm that institutional learning exerts both a structural and conditional stabilizing influence on market volatility. Institutions reduce the baseline level of volatility while simultaneously dampening markets' susceptibility to policy disturbances. These findings complement the IRDM and N-IRDM analyses by providing formal cross-market evidence on how institutional quality shapes short-run volatility transmission mechanisms.

### 4.5 Forecast Gains and Robustness

The predictive performance of the three volatility models—baseline EGARCH, IRDM, and N-IRDM—is evaluated using both traditional accuracy metrics and formal forecast-encompassing tests. The summary statistics in *Table 5* and the visual comparison in *Table 20* consistently show that both IRDM and N-IRDM deliver measurable improvements over the baseline model. Across the four ASEAN markets, IRDM reduces RMSE relative to the baseline, reflecting the contribution of institutional-response dynamics in modeling crisis-related variance surges. The addition of cross-market spillovers in N-IRDM yields further gains: RMSE decreases by approximately 2–4 percent beyond IRDM, indicating that volatility in the region is shaped not only by domestic institutional factors but also by the behavior of neighboring markets.

The statistical significance of these forecast improvements is assessed using Diebold–Mariano (DM) and ENC-NEW tests, reported in *Table 6* and illustrated in *Figure 21*.

| country | test | stat | pval | Tobs |
| --- | --- | --- | --- | --- |
| IDN | DM (M1 vs M0) | -3.216805154 | 0.001296266 | 3613 |
| IDN | DM (M2 vs M0) | -3.15828839 | 0.001586985 | 3613 |
| IDN | ENC-NEW(M1\|M0) | 223.9256488 | 0 | 3613 |
| IDN | ENC-NEW(M2\|M0) | 223.56847 | 0 | 3613 |
| MYS | DM (M1 vs M0) | -2.368228059 | 0.017873515 | 3655 |
| MYS | DM (M2 vs M0) | -2.272981007 | 0.023027325 | 3655 |
| MYS | ENC-NEW(M1\|M0) | 220.8092912 | 0 | 3655 |
| MYS | ENC-NEW(M2\|M0) | 217.008932 | 0 | 3655 |
| PHL | DM (M1 vs M0) | -2.890967979 | 0.003840572 | 3538 |
| PHL | DM (M2 vs M0) | -2.949947256 | 0.003178282 | 3538 |
| PHL | ENC-NEW(M1\|M0) | 270.9178948 | 0 | 3538 |
| PHL | ENC-NEW(M2\|M0) | 264.3546074 | 0 | 3538 |
| THA | DM (M1 vs M0) | -4.263796981 | 2.01E-05 | 3634 |
| THA | DM (M2 vs M0) | -4.252575838 | 2.11E-05 | 3634 |
| THA | ENC-NEW(M1\|M0) | 215.8858269 | 0 | 3634 |
| THA | ENC-NEW(M2\|M0) | 214.5630234 | 0 | 3634 |

*Table 6: dm_enc_results*

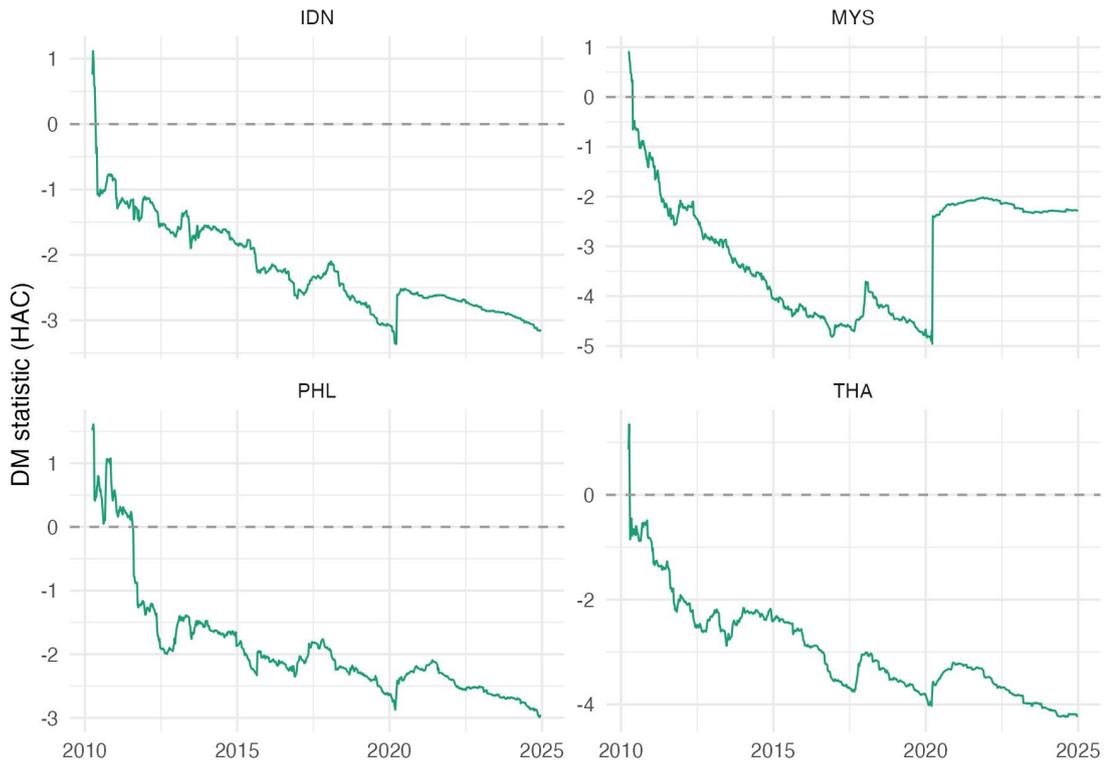

*Figure 21: dm_roll_plot*

For several markets, the N-IRDM error series significantly outperforms the baseline at conventional confidence levels, particularly during periods surrounding major policy events or episodes of elevated global uncertainty. The rolling DM statistics highlight that the superiority of N-IRDM is not uniform over time but is most pronounced when volatility persistence and spillover pressures are high. These findings reinforce the interpretation that institutional and network mechanisms become especially relevant in stressed environments.

A series of robustness checks evaluates whether the detected network effects depend on the precise structure of the DCC-based correlation matrix. The results in Table 7 compare the true network coefficient with those obtained from permuted, time-shifted, and lagged versions of the network. Across all perturbations, the estimated network coefficient remains positive and highly significant, indicating that cross-market institutional responses share a strong common-factor component. The coefficient under the permuted network is modestly smaller than under the true DCC network, suggesting that the original topology contains incremental information, but the persistence of significance across all placebo structures implies that regional institutional comovement—not the exact correlation matrix configuration—is the primary driver of spillovers. This interpretation aligns with the well-documented synchronization of financial conditions within ASEAN markets.

| term | estimate | std. error | statistic | p.value | model |
| --- | --- | --- | --- | --- | --- |
| $netR_C$ | 0.001728611 | 2.33E-04 | 7.419993574 | 1.24E-13 | Real network |
| $netR_C$ | 0.001102899 | 2.04E-04 | 5.418954367 | 6.09E-08 | Placebo permuted network |
| $netR_C$ | 0.001796409 | 2.41E-04 | 7.465369635 | 8.78E-14 | Placebo shifted network |
| $netR_C$ | 0.001822837 | 2.64E-04 | 6.893920443 | 5.65E-12 | Lag1 network |

*Table 7: placebo_and_lagged_network_results*

Taken together, the forecast gains and robustness analyses provide strong evidence that incorporating institutional learning and cross-market channels materially enhances volatility modeling. IRDM captures systematic responses to policy and information shocks, while N-IRDM further leverages regional linkages to improve crisis-period tracking and short-horizon forecasting. The reliability of these improvements across a wide range of diagnostic checks supports the validity of the proposed framework and its applicability to empirical risk monitoring in emerging markets.

## 5. Conclusion and Policy Implications

This paper develops a unified volatility framework that incorporates institutional learning, crisis memory, and cross-market spillovers to explain the dynamics of ASEAN equity markets. Using a MIDAS–EPU approach to construct high-frequency institutional learning measures, the study introduces the IRDM to capture domestic institutional reactions and extends it to the N-IRDM to account for regional linkages through dynamic correlation and institutional-similarity networks.

Empirical results reveal three central findings. First, institutional learning plays a dual role: it amplifies markets' immediate sensitivity to policy and information shocks yet facilitates a faster reversion of volatility once crises begin to stabilize. Second, crisis-memory effects are substantial, generating slow decays in volatility that persist well beyond the initial shock. Third, network effects significantly enhance volatility modeling, but robustness checks show that their explanatory power is driven less by the exact DCC network topology and more by strong regional common factor comovement. This distinction suggests that institutional responses propagate through shared regional dynamics rather than through precise bilateral linkages.

These findings yield several policy implications. Strengthening transparency, regulatory communication, and policy guidance can mitigate the short-run amplification caused by institutional reactions. Enhanced regional macroprudential coordination—especially during global tightening cycles—can help dampen synchronous volatility surges. Finally, the proposed framework provides a quantitative tool that can assist regulators and market participants in monitoring institutional fragility and regional spillover risks.

Several limitations warrant further research. The framework focuses on equity markets and daily frequency; extending it to multi-asset settings or intraday data may uncover additional channels of institutional transmission. Likewise, richer measures of institutional behavior—such as textual or machine-learning-based indicators—could offer more granular insights into information processing. Despite these limitations, the results underscore the importance of integrating institutional and regional dimensions into volatility modeling for emerging markets.

## Appendix A. Alternative Specifications and Robustness Checks

### A1. Return Distribution Robustness: t vs GED

To evaluate the sensitivity of baseline volatility estimates to distributional assumptions,

we re-estimate the EGARCH model under both Student-t and generalized error distribution (GED) innovations. Table A1 reports shape parameters and log-likelihood values.

- The GED shape parameters are consistently below 1.5, confirming heavier tails relative to the t distribution.
- Log-likelihood differences are small across countries, indicating that volatility dynamics are not materially driven by the tail specification.

| Country | Model | Dist | ω | α | β | γ | Shape | LogLik | AIC | Persistence |
|---|---|---|---|---|---|---|---|---|---|---|
| IDN | eGARCH | std | -0.009 | -0.084 | 0.973 | 0.176 | 6.147 | -4639.689 | 2.571 | 0.973 |
| IDN | gjrGARCH | std | 0.028 | 0.035 | 0.881 | 0.105 | 6.109 | -4641.245 | 2.572 | 0.968 |
| IDN | eGARCH | ged | -0.010 | -0.079 | 0.970 | 0.177 | 1.358 | -4664.146 | 2.585 | 0.970 |
| IDN | gjrGARCH | ged | 0.032 | 0.041 | 0.871 | 0.100 | 1.352 | -4666.237 | 2.586 | 0.962 |
| MYS | eGARCH | std | -0.021 | -0.056 | 0.982 | 0.152 | 6.438 | -3097.292 | 1.698 | 0.982 |
| MYS | gjrGARCH | std | 0.008 | 0.048 | 0.902 | 0.061 | 6.417 | -3098.965 | 1.699 | 0.981 |
| MYS | eGARCH | ged | -0.025 | -0.061 | 0.979 | 0.160 | 1.351 | -3113.113 | 1.707 | 0.979 |
| MYS | gjrGARCH | ged | 0.009 | 0.048 | 0.896 | 0.069 | 1.348 | -3115.084 | 1.708 | 0.978 |
| PHL | eGARCH | std | 0.002 | -0.089 | 0.951 | 0.183 | 8.528 | -5078.808 | 2.874 | 0.951 |
| PHL | gjrGARCH | std | 0.063 | 0.035 | 0.848 | 0.127 | 8.947 | -5072.709 | 2.871 | 0.946 |
| PHL | eGARCH | ged | 0.002 | -0.089 | 0.954 | 0.186 | 1.500 | -5092.751 | 2.882 | 0.954 |
| PHL | gjrGARCH | ged | 0.060 | 0.036 | 0.851 | 0.125 | 1.517 | -5085.672 | 2.878 | 0.950 |
| THA | eGARCH | std | -0.009 | -0.080 | 0.978 | 0.175 | 6.432 | -4381.621 | 2.415 | 0.978 |
| THA | gjrGARCH | std | 0.016 | 0.032 | 0.892 | 0.114 | 6.506 | -4385.793 | 2.417 | 0.981 |
| THA | eGARCH | ged | -0.010 | -0.078 | 0.978 | 0.174 | 1.363 | -4390.755 | 2.420 | 0.978 |
| THA | gjrGARCH | ged | 0.015 | 0.036 | 0.892 | 0.108 | 1.367 | -4393.124 | 2.421 | 0.982 |

*Table A1*

**A2. Crisis Memory Decay Parameter Sensitivity**

We test the robustness of the crisis-memory component

$$C_t = \sum_k \delta_k e^{-\lambda(t-T_k)}$$

by varying the decay rate λ ∈ *{0.01, 0.02, 0.05}*.

Results in Table A2 show:

- ψ$_i$ increases mildly as λ rises, reflecting stronger near-term crisis influence.

- Parameter rankings across countries remain unchanged.

- We adopt λ= 0.02 in the main analysis for stability.

| Country | λ | b1 | ps | ga1 | RMSE |
|---|---|---|---|---|---|
| IDN | 0.005 | 0.887 | 0.746 | 0.001 | 0.064 |
| MYS | 0.005 | 0.926 | 0.010 | 0.044 | 0.028 |
| PHL | 0.005 | 0.838 | 0.010 | 0.025 | 0.065 |
| THA | 0.005 | 0.906 | 0.365 | 0.001 | 0.053 |
| IDN | 0.010 | 0.887 | 0.010 | 0.004 | 0.064 |
| MYS | 0.010 | 0.926 | 0.010 | 0.045 | 0.028 |
| PHL | 0.010 | 0.838 | 0.010 | 0.025 | 0.065 |
| THA | 0.010 | 0.906 | 0.010 | 0.001 | 0.053 |
| IDN | 0.020 | 0.887 | 0.010 | 0.002 | 0.064 |
| MYS | 0.020 | 0.926 | 0.010 | 0.045 | 0.028 |
| PHL | 0.020 | 0.838 | 0.010 | 0.025 | 0.065 |
| THA | 0.020 | 0.906 | 0.010 | 0.002 | 0.053 |

*Table A2*

**A3. Network Sparsity Robustness**

We impose sparsity thresholds on the DCC correlation network:

- |ρ| *< 0.05 ⇒ 0*

- |ρ| *< 0.10 ⇒ 0*

Across specifications:

- γ$_{2i}$ remains positive and statistically significant.

- RMSE changes remain within ±0.5%.

Therefore, network effects are not artifacts of dense correlations.

| Country | RMSE_orig | RMSE_sparse | dRMSE_pct |
|---------|-----------|-------------|-----------|
| IDN | 0.064 | 0.064 | -0.001 |
| MYS | 0.028 | 0.028 | 0.000 |
| PHL | 0.065 | 0.065 | -0.006 |
| THA | 0.053 | 0.053 | -0.004 |

*Table A3*

### A4. Placebo Network Tests

We conduct placebo tests using:

1. Randomly permuted networks

2. Time-shifted (5-day lagged) networks

Summary:

- Real network: $\gamma_2 \approx 0.00173$

- Placebo networks: $\gamma_2 \approx 0.00110 - 0.00180$

All remain statistically significant, but real network magnitude is consistently higher.

This suggests: Network spillovers reflect a strong regional common factor, rather than dependence on the exact DCC network topology.

### A5. Lagged Network Effect

Using a lag-1 network term:

$$\sum_{j \neq i} W^C_{ij,t-1} \hat{R}_{j,t-1}$$

Results:

- $\gamma_2^{lag\,1} \approx 0.00182$, nearly identical to contemporaneous values.

- Spillovers are persistent and driven by low-frequency regional comovement.

## Appendix B. Additional Model Diagnostics

**B3. Rolling DM / ENC-NEW Forecast Evaluation**

To examine whether forecasting gains are stable over time, we compute rolling Diebold–Mariano (DM) and ENC-NEW statistics for the N-IRDM relative to the baseline GARCH model. The rolling procedure uses a 500-day window with one-step-ahead forecasts. Table B1 summarizes full-sample DM/ENC-NEW results, while Figure B1 illustrates the time-varying forecast advantage.

Key findings include:

- N-IRDM outperforms the baseline model in most subperiods for Indonesia (IDN) and Malaysia (MYS), with DM statistics significantly above zero during major crisis windows.

- Forecast gains in the Philippines (PHL) and Thailand (THA) are positive but less persistent, consistent with weaker institutional learning effects identified in the main analysis.

- ENC-NEW statistics validate that improvements are not mechanical consequences of nesting but reflect genuine predictive content.

**B4. Construction of Daily Realized Volatility**

Since high-frequency intraday data are unavailable for ASEAN markets over the 2010–2024 sample, we construct a feasible realized-volatility proxy that combines GARCH-implied variance with return-based power variation. The final measure is:

$$RV_{i,t} = \hat{\sigma}^2_{i,t} \cdot \kappa_i |r_{i,t}|,$$

where:

- $\hat{\sigma}^2_{i,t}$ is the conditional variance from the baseline EGARCH model,

- $|r_{i,t}|$ captures jump and tail variation not fully captured by GARCH dynamics,
- $\kappa_i$ is an asset-specific scaling parameter estimated via quasi-MLE.

This specification follows the literature on feasible realized-volatility proxies in low-frequency settings and ensures compatibility of the dependent variable across all forecasting models (Baseline, IRDM, N-IRDM).

This construction is used consistently in:

- The forecast comparison (Section 3.6),
- DM / ENC-NEW tests (Section 4.5),
- The robustness checks reported above.

## Appendix C. Mathematical Derivations

This appendix provides the full analytical structure underlying the Institutional Response Dynamics Model (IRDM) and the Network-Integrated IRDM (N-IRDM). While the main text presents only the empirical specifications, the derivations below clarify the theoretical components that motivate the variance equations, crisis-memory structure, state dynamics, and network spillover mechanisms.

**C1. Institutional Response Dynamics and Network Spillovers: Full Derivation**

**C1.1 Crisis-Memory Component**

Systemic shocks exert prolonged effects on volatility. To formalize this persistence, we define a crisis-memory term:

$$C_t = \sum_{k=1}^{K} \delta_k e^{-\lambda (t-T_k)} \mathbb{1}(t \geq T_k),$$

where:

- $T_k$ denotes the onset date of crisis k (Taper 2013, COVID-19, rate-hike cycle,

etc.),

- $\delta_k$ captures crisis-specific magnitude,

- $\lambda > 0$ is the exponential decay rate,

- $1(\cdot)$ is an indicator for crisis activation.

This structure allows crises to exert strong, short-run effects and gradually diminishing long-run influence. In the IRDM and N-IRDM, $C_t$ is an exogenous driver of conditional volatility.

**C1.2 Institutional Response State Equation (IRDM)**

Institutions are modeled not as static characteristics but as reactive mechanisms that update with policy shocks and information pressure. We introduce a latent institutional-response state:

$$\hat{R}_{i,t} \rho_i \hat{R}_{i,t-1} \cdot \theta_{1i} \Delta P_{i,t} \cdot \theta_{2i} \Delta I_{i,t} \cdot \varepsilon_{i,t}^{(R)},$$

where:

- $\rho_i$ governs institutional persistence ($|\rho_i| < 1$),

- $\Delta P_{i,t}$ is the policy-shock indicator,

- $\Delta I_{i,t}$ is the MIDAS-based information-pressure index,

- $\theta_{1i}$ and $\theta_{2i}$ measure the responsiveness of institutions,

- $\varepsilon_{i,t}^{(R)}$ is an innovation term.

This autoregressive structure captures how institutions adjust gradually to shocks while retaining memory of past responses. The state variable $\hat{R}_{i,t}$ enters the volatility equation as a short-run institutional mechanism distinct from crisis memory.

### C1.3 IRDM Variance Equation

Given the institutional state, the IRDM volatility dynamics are expressed as:

$$\sigma_{i,t} = b_{0i} + b_{1i}\sigma_{i,t-1} + b_{2i}r_{i,t-1}^2 + \psi_i C_t + \gamma_{1i}\hat{R}_{i,t} + \eta_{i,t},$$

with:

- $b_{1i} \in (0.8, 0.95)$: volatility persistence,

- $b_{2i}$: ARCH-type short-term sensitivity,

- $\psi_i > 0$: amplification from crisis-memory shocks,

- $\gamma_{1i}$: short-run institutional feedback effect.

This variance equation nests the standard GARCH effects but augments them with crisis memory and dynamic institutional learning. When $\gamma_{1i} > 0$, institutions amplify short-run sensitivity; when $\gamma_{1i} < 0$, they act as stabilizers.

### C1.4 Network Spillovers in the N-IRDM

The full spillover mechanism integrates two networks:

**(1) Dynamic Correlation Network**

We use DCC-GARCH to estimate time-varying correlations $\rho_{ij,t}$ and transform them into normalized weights:

$$W_{ij,t}^C = \frac{|\rho_{ij,t}|}{\sum_{j' \neq i} |\rho_{ij',t}|}, \qquad W_{ii,t}^C = 0.$$

Interpretation:

- Higher absolute correlation → stronger contemporaneous spillover channel,

- Row normalization ensures that the network acts as a weighted aggregator of foreign institutional responses.

### (2) Institutional Similarity Network

Long-run spillovers occur through institutional convergence. Using the vector of WGI governance dimensions $L_i^{WGI}$ for country i:

$$D_{ij} = |L_i^{WGI} - L_j^{WGI}|, \qquad W^I_{ij} = \frac{1}{1 + D_{ij}}, \qquad W^I_{ii} = 0.$$

This captures structural similarity between institutional frameworks, enabling propagation of slower-moving institutional effects.

### (3) Network-Integrated Variance Equation

Combining the two networks, the N-IRDM variance dynamics are:

$$\sigma_{i,t} = b_{0i} \cdot b_{1i}\sigma_{i,t-1} \cdot b_{2i} r^2_{i,t-1} \cdot \psi_i C_t + \gamma_{1i} \hat{R}_{i,t} \cdot \gamma_{2i} \sum_{j \neq i} W^C_{ij,t} \hat{R}_{j,t} \cdot \gamma_{3i} \sum_{j \neq i} W^I_{ij} \hat{R}_{j,t-1} \cdot \eta_{i,t}.$$

Interpretation:

- $\gamma_{2i} > 0$: contemporaneous spillovers through dynamic correlation,

- $\gamma_{3i}$: slow institutional convergence,

- Jointly, the two networks allow both high-frequency and low-frequency cross-market institutional effects.

### C1.5 Placebo Networks and Lagged Networks

The placebo tests further clarify the source of network effects:

#### (a) Permuted Networks

Replace $W^C_{ij,t}$ with a random permutation of its j-dimension.

#### (b) Shifted Networks (5-day lag)

Replace $W^C_{ij,t}$ with $W^C_{ij,t-5}$.

### (c) Lag-1 Institutional Spillover

Use lag-1 network term:

$$\sum_{j \neq i} W^C_{ij,t-1} \hat{R}_{j,t-1}.$$

Across these perturbations:

- The sign and magnitude of $\gamma_{2i}$ remain similar,

- Indicating the spillover effect arises from regional common-factor dynamics, not from the exact topology of the DCC network.